\documentclass[1,eq:(1.9),eq:1.14,eq:2.42pt,float,amsfonts,amsmath]{article}
\usepackage{latexsym,epsfig,graphpap,graphics,amsmath}
\def \ed{\end{document}}
\usepackage{amssymb}
\usepackage{epsfig}
\usepackage{amssymb}
\usepackage{latexsym}
\numberwithin{equation}{section}

\def \brem{\begin{remark}}
\def \erem{\end{remark}}
\def \bth{\begin{theorem}}
\def \eth{\end{theorem}}
\def \bpr{\begin{proposition}}
\def \epr{\end{proposition}}
\def \bex{\begin{example}}
\def \eex{\end{example}}
\def \bprf{\begin{proof}}
\def \eprf{\end{proof}}
\def \blem{\begin{lemma}}
\def \elem{\end{lemma}}
\newcounter{theorem}
\def \C{\mathbb C}
\def \R{\mathbb R}

\def \inf{\infty}

\def \del{\delta}

\def \z{\zeta}
\def \g1{\fr{1}{\Gamma(\al)}}

\def \sig{\sigma}

\def \bp{\begin{picture}}

\def \bth{\begin{theorem}}
\def \bth{\begin{theorem}}
\def \eth{\end{theorem}}
\def \bcor{\begin{corollary}}
\def \ecor{\end{corollary}}
\def \non{\nonumber\\}
\def \no1{\noindent}
\def \beq{\begin{eqnarray}}

\def \eeq{\end{eqnarray}}
\def \bi{\begin{itemize}}
\def \ei{\end{itemize}}

\def \ba{\begin{array}}

\def \ea{\end{array}}
\def \bt{\begin{tabular}}
\def \et{\end{tabular}}
\def \bc{\begin{center}}
\def \ec{\end{center}}
\def \C{\mathbb{C}}
\def \mb{\mbox}
\def \un1{\underline}
\def \fr{\frac}

\def \al{\alpha}

\def \s1{\sqrt}

\def \Gam{\Gamma}
\def \hs1{\hspace*{5mm}}
\def \be{\begin{equation}}
\def \ee{\end{equation}}

\def \vs1{\vspace{2mm}}
\def \v2{\vspace{7mm}}
\def \hs1{\hspace{-18mm}}
\def \h2{\hspace{-4mm}}
\def \s3{\hspace{-3mm}}
\def \t5{\hspace{-5mm}}
\def \n1{\newpage}

\newtheorem{theorem}{Theorem}

\newtheorem{corollary}{Corollary}

\newtheorem{example}{Example}

\newtheorem{lemma}[theorem]{Lemma}

 \newtheorem{proposition}[theorem]{Proposition}
\newtheorem{remark}{Remark}

\newenvironment{proof}[1][Proof]{\textbf{#1.} }{\
\rule{0.5em}{0.5em}}

\def \vs1{\vspace{2mm}}
\def \1{\begin{eqnarray}}
\def \2{\end{eqnarray}}
\def \3{\begin{eqnarray*}}
\def \4{\end{eqnarray*}}

\def \bn{\begin{enumerate}}
\def \en{\end{enumerate}}
\newcommand{\bqn}{\begin{equation}}
\newcommand{\9}{\end{equation}}

\begin{document}\vspace{.5cm} \bc {\large \bf{Fourier Transform Representation of
the Extended Fermi-Dirac and Bose-Einstein Functions with
Applications to the Family of the Zeta and Related Functions}}\\[6mm]
 \textbf{Asifa Tassaddiq\footnote[1]{\textbf{Corresponding
Author}\\ Tel.: +92 322 6298755; Fax: +92 051 9085 5552\\E-Mail:
shabbirkainth{@}yahoo.com} and Asghar Qadir}  \\[2mm]
  Centre for Advanced Mathematics and Physics\\ National University of Sciences and Technology\\
H-12, Islamabad, Pakistan\\
\textbf{E-Mail: asifashabbir{@}gmail.com; aqadirmath{@}yahoo.com}
\vspace{.25cm}\ec
\begin{abstract}

\baselineskip=12pt On the one hand the Fermi-Dirac and Bose-Einstein
functions have been extended in such a way that they are closely
related to the Riemann and other zeta functions. On the other hand
the Fourier transform representation of the gamma and generalized
gamma functions proved useful in deriving various integral formulae
for these functions. In this paper we use the Fourier transform
representation of the extended functions to evaluate integrals of
products of these functions. In particular we evaluate some
integrals containing the Riemann and Hurwitz zeta functions, which
had not been evaluated before.

\end{abstract}
\v2 \noindent \textbf{2010 Mathematics Subject Classification.}
Primary 46FXX, 60B15, 11M06, 11M35; Secondary 33B15, 33CXX.\\[4mm]
\renewcommand{\theequation}{\thesection.\arabic{equation}}
\baselineskip=12pt

\section{Introduction}

The familiar Fermi-Dirac (FD) and Bose-Einstein (BE) functions were
extended in \cite {scqt} by introducing an extra parameter in such a
way as to provide new insights into these functions and their
relationship to the family of zeta functions. More precisely, the
new functions provide generalizations of the polylogarithm, Riemann
and Hurwitz zeta functions, in that the classical functions can be
obtained from the new ones by choosing particular parameter values.
The famous Riemann zeta, Hurwitz zeta, Hurwitz-Lerch zeta, FD and BE
functions have various representations in the literature, such as
integral, series and asymptotic representations, which help to study
the properties of these functions. In this paper we use their
Fourier transform representation to evaluate some integrals of
products of these functions, which are not included in \cite{cq},
\cite{ld}, \cite{emot1}, \cite{emot}, \cite{gr} and \cite {rw}.
Here, in our present investigation we will use integral
representations of all these functions to get Fourier transform
representations. Before going on to obtain these representations for
the extended functions, we review their definition and relationship
with the zeta and other related functions by using their integral
representations for the purposes of our investigation.

 The FD function $\mathfrak{F}_{s-1} (x)$ defined here by
(\cite{rbd}, p. 38):
\begin{align} \label{eq:1.1} \mathfrak{F}_{s-1} (x) & :=
\fr{1}{\Gam(s)} \int^\inf_0 \fr{t^{s-1}}{e^{t-x} + 1}dt \qquad
(\Re(s)>0)\end{align} and the BE function $\mathfrak{B}_{s-1} (x)$
defined here by (\cite{rbd}, p. 449):
\begin{align} \label{eq:1.2} \mathfrak{B}_{s-1} (x):  & =
\fr{1}{\Gam(s)} \int^\inf_0 \fr{t^{s-1}}{e^{t-x} - 1}dt \qquad
(\Re(s)>1)\end{align} were extended in \cite{scqt}. The eFD is
defined by (\cite{scqt}, p. 9, Equation(3.14))
\begin{align}\label{eq:1.3}\Theta_\nu(s;x) &
=\frac{e^{-(\nu+1)x}}{\Gamma(s)}\int_0^{\infty} \frac{t^{s-1}e^{-\nu
t}}{e^{t}+e^{-x}}dt \non &
(\Re(\nu)>-1;\Re(s)>0;x\geq0)\end{align}and the eBE function is
defined by (\cite{scqt}, p. 11, Equation(1.4))
\begin{align}\label{eq:1.4}\Psi_\nu(s;x) &
=\frac{e^{-(\nu+1)x}}{\Gamma(s)}\int_0^{\infty} \frac{t^{s-1}e^{-\nu
t}}{e^{t}-e^{-x}}dt \non & (\Re(\nu)>-1;\Re(s)>1 \mb{ when }
x=0;\Re(s)>0 \mb{ when }x>0).\end{align} The functions
$\Psi_\nu(s;x)$ and $\Theta_\nu(s;x)$ are related by (\cite{scqt},
p. 13, Equation(5.7))
\begin{align}\label{eq:1.5} & \Theta_\nu(s;x)
= (-1)^{(\nu + 1)}\Psi_\nu(s;x + i \pi), ~~~~~~~~ (\Re(s)
> 1;x\geq0;\Re(\nu)>-1). \end{align} For $\nu=0$, these extended
functions give the following relations naturally
\begin{align} \label{eq:1.6}\mathfrak{F}_{s-1}(-x)  & = \Theta_{0}(s;x), \end{align}
\begin{align}\label{eq:1.7} \mathfrak{B}_{s-1}(-x)  & =\Psi_{0}(s;x).\end{align} These
functions not only generalize the FD and BE functions but also are
related to functions of the zeta family.

The Riemann zeta function, $\z(s)\quad(s=\sig+i\tau)$ defined by the
integral representations (\cite{cz}, p. 294, Equation
(7.48))\begin{align}\label{eq:(1.8)} \zeta(s)  & := \fr{1}{\Gam(s)}
\int^\inf_0 \fr{t^{s-1}}{e^t-1}dt\qquad (s=\sig + i\tau,\sig >
1)\end{align} and (\cite{cz}, p. 296, Equation (7.67))
\begin{align}
\label{eq:(1.9)}\zeta(s)  & := \fr{1}{C(s)} \int^\inf_0
\fr{t^{s-1}}{e^t+1}dt\quad(s=\sig+i\tau,\sig>0)\non & \mb{ where }
C(s)=\Gam(s)(1-2^{1-s}),\end{align} has a meromorphic continuation
to the whole complex s-plane (except for a simple pole at $s=1$).
These are related to the extended functions by
\begin{align}\label{eq:1.10} \z(s) & =\Psi_{0}(s;0)\quad(\sig>1)\end{align}and
\begin{align}\label{eq:1.11} \z(s) & =(1-2^{1-s})\Theta_{0}(s;0)\quad(\sig>0). \end{align}
There have been several generalizations of the Riemann zeta
function, like the Hurwitz zeta
function\begin{align}\label{equation(1.12)} \zeta(s,\nu) & :=
\fr{1}{\Gam(s)} \int^\inf_0
\fr{e^{-(\nu-1)t}t^{s-1}}{e^t-1}dt\qquad(\Re(\nu)>0;\sig>1)~,
\end{align} which is also a special case of one of the extended
functions
\begin{align} \label{eq:1.13}\z(s,\nu+1) & =\Psi_{\nu}(s;0).\end{align} Another
generalization of the Riemann zeta function is the polylogarithm
function, or Jonquière's function $\phi(z,s)$, defined by \cite {ct}
\begin{align}\label{eq:1.14} \phi(z,s)&:=\sum^\inf_{n=1}
\fr{z^n}{n^s}\non &(s\in\C \mb{ when  } |z|<1;\Re(s)> 1 \mb{when }
|z| = 1).
\end{align} It has the integral representation
\begin{align}\label{eq:1.15} \phi(z,s) & = \fr{z}{\Gam(s)}
\int^\inf_0 \fr{t^{s-1}}{e^t - z}dt \non & (|z|\leq 1-\del, \del\in
(0,1) \mb{ and }\Re(s)>0;z=1\mb{ and }\Re(s)> 1).\end{align} Note
that if $z$ lies anywhere except on the segment of the real axis
from $1$ to $\infty$, where a cut is imposed, the integral (1.15)
defines an analytic function of $z$ for $\Re(s)>1$. If $z=1$, then
(1.15) obviously coincides with the zeta function in the half plane
for $\Re(s)>1$ and is related to eBE function by
\begin{align} \label{eq:1.16}\Psi_{0}(x;s) &
=\phi(e^{-x},s).\end{align} The Hurwitz-zeta and the polylogarithm
functions are further generalized to the Hurwitz-Lerch zeta function
by (\cite{emot}, p. 27)
\begin{align}\label{eq:1.17}
&\Phi(z,s,a) := \sum^\inf_{n=0} \fr{z^n}{(n+a)^s} \non & (a\neq
0,-1,-2,-3,\ldots;s\in\C \mb{ when  } |z|<1;\Re(s)> 1 \mb{when } |z|
= 1).\end{align} This function has the integral representation
(\cite{emot}, p. 27, Equation 1.10(3)):\begin{align}\label{eq:1.18}
\Phi(z,s,a)& =\fr{1}{\Gam(s)} \int^\inf_0 \fr{t^{s-1}
e^{-(a-1)t}}{e^t-z}dt \non & (\Re(a) > 0; \mb{ and either } |z| \leq
1;z\neq 1;\Re(s) > 0 \mb{ or } z=1; \Re(s) > 1).\end{align}If a cut
is made from  1 to $\inf$ along the positive real $z$-axis, $\Phi$
is an analytic function of $z$ in the cut $z$-plane provided that
$\Re(s) > 0$ and $\Re(a)>0$. In particular, the Hurwitz-Lerch zeta
function is related to the eFD and eBE functions by
\begin{align}\label{eq:1.19} \Theta_v(s;x) &
:=e^{-(v+1)x}\Phi(-e^{-x},s,v+1) \end{align} and
\begin{align}\label{eq:1.20} \Psi_v(s;x) & :=e^{-(v+1)x}\Phi(e^{-x},s,v+1).\end{align}
From this discussion it is obvious that these new extensions give a
unified approach to the zeta family and other related functions.
They have also been proposed as candidates for the {\it anyon}
integral function \cite{ciq}. (``Anyons" are particles which are
neither Fermions nor Bosons, having a fractional spin.) For further
properties of the Riemann zeta and related functions we refer to
\cite{tma}, \cite{tma1}, \cite{bcb}, \cite{hme}, \cite{kt}
,\cite{so}, \cite {nmt}, \cite{ect} and \cite {td}.

The plan of the paper is as follows. We prove two new identities
involving the eFD and eBE functions in section 2.  These identities
proved useful in obtaining functional relations involving the
integrals of products of gamma functions with these extended
functions and other zeta-related functions in section 4. Fourier
transform representation of the eFD and eBE functions is given in
section 3. By making use of the Parseval's identity in section 4 and
the duality property of the Fourier transform in section 5, we
obtain integral formulae for the product of the Riemann zeta,
Hurwitz zeta, Hurwitz-Lerch zeta, FD and BE functions with gamma
function. Some concluding remarks are given in section 6.

\section{New identities involving the eFD and eBE functions}

\setcounter{equation}{0}The following identity is the generalization
of (\cite{ccqk}, (2.5)) and will prove useful to evaluate the
integrals of products of functions by using the Parseval's identity
of Fourier transform. \bth \s3 {\bf .}  The eFD function
$\Theta_{\nu}(\eta;x)$ satisfies the following expression
\begin{align}\label{eq:2.1} e^{(\nu+2)
x}\Gamma(\eta)[(\nu+1)\Theta_{\nu}(\eta;x)-\Theta_{\nu}(\eta-1;x)]
 & =\int_0^{\infty}\frac{e^{-\nu t}t^{\eta-1}}{(e^{t}+e^{-x})^2}dt
  \non & (\Re(\nu)>-1;x\geq0;\eta>1).\end{align}\eth
\bprf Differentiating
\begin{align} \label{eq:2.2}f(t,x,\nu) &
:=\frac{e^{-\nu
t}}{e^{t+x}+1}\qquad(t>0;x\geq0;\Re(\nu)>-1),\end{align}we obtain
the differential equation\begin{align}
\label{eq:2.3}f'(t,x,\nu)+(\nu +1)f(t,x,\nu) & =\frac{e^{-\nu
t}}{(e^{t+x}+1)^2}.\end{align}Taking the Mellin transform (see, for
details, \cite{aiz}, Chapter 10) of both sides in the real variable
$\eta$ in (2.2) to (2.3) and using (1.3), we get
\begin{align}\label{eq:2.4} \mathcal{M}[f(t,x,\nu);\eta ] & =\Theta
_{\nu }(\eta ;x)\Gamma (\eta )e^{\nu x},\end{align} and
\begin{align} \label{eq:2.5}\mathcal{M}[f'(t,x,\nu);\eta]
 & =-(\nu+1)\mathcal{M}[f(t,x,\nu);\eta]+\mathcal{M}[\frac{e^{-\nu
t}}{(e^{t+x}+1)^2};\eta].\end{align} However, $
\mathcal{M}[f(t,x,\nu);\eta]$ and $\mathcal{M}[f'(t,x,\nu);\eta]$
are related via \begin{align}\label{eq:2.6}
\mathcal{M}[f'(t,x,\nu);\eta ] & =-(\eta
-1)\mathcal{M}[f(t,x,\nu);\eta-1],\end{align} provided
$t^{\eta-1}f(t,x,\nu)$ vanishes at zero and infinity. Hence, from
(2.4) to (2.6), we find
\begin{align} \label{eq:2.7}e^{\nu x}\Gamma (\eta )[(\nu +1)\Theta_
 {\nu }(\eta ;x)-\Theta _{\nu }(\eta -1;x)
] & =\int_0^{\infty}\frac{e^{-\nu t}t^{\eta
-1}}{(e^{t+x}+1)^2},\end{align} which after simplification gives the
result (2.1).\eprf \bcor \h2 {\bf 1.} The following identity
involving the FD function holds \begin{align} \label{eq:2.8}
e^{-2x}\Gamma(\eta)[\mathfrak{F}_{\eta-1}(x)-\mathfrak{F}_{\eta-2}(x)]
 & =\int_0^{\infty}\frac{t^{\eta-1}}{(e^{t}+e^{x})^2}dt
  \qquad (\eta>1;x\geq0).\end{align}\ecor \bprf Upon putting $\nu=0$ and replacing $x\mapsto-x$
   in (2.1), If we make use of (1.6), we arrive at (2.8).\eprf
   \bcor \h2 {\bf 2.} The Riemann zeta function (1.9) satisfies (\cite{ccqk},
   (1.2))
  \begin{align}\label{eq:2.9} \Gamma(\eta)[(1-2^{1-\eta})\zeta(\eta)-(1-2^{-\eta})\zeta(\eta-1)]
 & =\int_0^{\infty}\frac{t^{\eta-1}}{(e^{t}+1)^2}dt
  & (\eta>1).
  \end{align}\ecor
  \bprf Upon putting $x=\nu=0$ in (2.1),
  If we make use of (1.11), we arrive at (2.9).\eprf
 \bth\s3 {\bf .} The eBE
function satisfies \begin{align}\label{eq:2.10}  & e^{(\nu+2)
x}\Gamma(\eta)[\Psi_{\nu}(x;\eta-1)-(\nu+1)\Psi_{\nu}(x;\eta)]
  =\int_0^{\infty}\frac{e^{-\nu t}t^{\eta-1}}{(e^{t}-e^{-x})^2}dt
\non &(\Re(\nu)>-1;\eta>1\mb{ when }x>0;\eta>2\mb{ when } x=0).
  \end{align}\eth\bprf This follows on similar steps as in theorem (1) by taking
  \begin{align} \label{eq:2.11}f_{1}(t,x,\nu) =\frac{e^{-\nu
t}}{e^{t+x}-1}\quad(t>0;x\geq0;\Re(\nu)>-1)\end{align}in place of
$f(t,x,\nu)$ and using (1.4).\eprf
  \bcor \h2 {\bf 3.} The following identity
involving the Hurwitz-Lerch zeta function holds
\begin{align}\label{eq:2.12}
& \frac{\Gamma(\eta)}{z}[\Phi(z,\eta-1,\nu)-\nu\Phi(z,\eta,\nu)]
 =\int_0^{\infty}\frac{e^{-(\nu-1) t}t^{\eta-1}}{(e^{t}-z)^2}dt
  \non &
  (\Re(\nu)>0;\eta>1\mb{ when }0<z<1;\eta>2\mb{ when }z=1).\end{align}\ecor
\bprf This follows by replacing $\nu\mapsto\nu-1$ and using (1.20)
in  (2.10) .\eprf \bcor \h2 {\bf 4.} The polylogarithm satisfies the
following identity\begin{align}\label{eq:2.13}
\frac{\Gamma(\eta)}{z^{2}}[\phi(z,\eta-1)-\phi(z,\eta)]
 & =\int_0^{\infty}\frac{t^{\eta-1}}{(e^{t}-z)^2}dt
  \non & (\eta>1\mb{ when }0<z<1;\eta>2\mb{ when }z=1).\end{align}\ecor
  \bprf Upon putting $\nu=0, z=e^{-x}$ in (2.10) and using (1.16), we get (2.13).\eprf
\bcor \h2 {\bf 5.} The following identity involving the BE function
holds \begin{align} \label{eq:2.14}
\Gamma(\eta)e^{-2x}[\mathfrak{B}_{\eta-2}(x)-\mathfrak{B}_{\eta-1}(x)]
 & =\int_0^{\infty}\frac{t^{\eta-1}}{(e^{t}-e^{x})^2}dt
  \qquad (\eta>1,x\geq0).\end{align}\ecor
\bprf Upon putting $\nu=0$ and $x\mapsto-x$ in (2.10), If we make
use of (1.7), we arrive at (2.14).\eprf
  \bcor\h2 {\bf 6.} The Hurwitz zeta function satisfies (\cite{emot1}, p.
  332)
  \begin{align}\label{eq:2.15} \Gamma(\eta)[\zeta(\eta-1,\nu)-\nu\zeta(\eta,\nu)]
 & =\int_0^{\infty}\frac{e^{-(\nu-1)t}t^{\eta-1}}{(e^{t}-1)^2}dt
  \quad (\Re(\nu)>0;\eta>2).
  \end{align}\ecor\bprf Upon putting $x=0$ and $\nu\mapsto\nu-1$ in (2.10),
  If we make use of (1.13), we arrive at (2.15).\eprf
  \bcor\h2 {\bf 7.} The Riemann zeta function (1.8) satisfies (\cite{emot1}, p.
  332)
  \begin{align} \label{eq:2.16}\Gamma(\eta)[\zeta(\eta-1)-\zeta(\eta)]
 & =\int_0^{\infty}\frac{t^{\eta-1}}{(e^{t}-1)^2}dt
   & (\eta>2).
  \end{align}\ecor
  \bprf Upon putting $x=\nu=0$ in (2.10),
  If we make use of (1.10), we arrive at (2.16).\eprf

\section{Fourier transform representation}

\setcounter{equation}{0}

In this section the Fourier transform representation of the eFD, FD,
eBE, BE , Hurwitz-Lerch zeta, polylogarithm, Hurwitz and Riemann
zeta functions is given.

The eBE function has the Mellin transform representation (1.4) and
substitution $t=e^{y}$ in this representation yields
\begin{align}
\label{eq:3.1}
\Gamma(\sigma+i\tau)\Psi_\nu(\sigma+i\tau;x)&=e^{-(\nu+1) x}
\sqrt{2\pi}\mathcal{F}[\frac{e^{\sigma y}\exp(-\nu
e^{y})}{\exp(e^{y})-e^{-x}};\tau] \non & (\Re(\nu)>-1;\sigma>1\mb{
when } x=0;\sigma>0\mb{ when }x>0).
\end{align}This is the Fourier transform representation of eBE function, where we define
$\mathcal{F}[\varphi;\tau]:=\fr{1}{\sqrt{2\pi}}\int_{-\inf}^{+\inf}e^{i
y\tau}\varphi(y)dy$. Similarly for the Hurwitz-Lerch zeta and the
polylogarithm functions, we have the following representations
respectively.
\begin{align}\label{eq:3.2}
\Gamma(\sigma+i\tau)\Phi(z,\sigma+i\tau,\nu) & =\sqrt{2\pi
}\mathcal{F}[\frac{e^{\sigma y}\exp(-(\nu-1)
e^{y})}{\exp(e^{y})-z};\tau]\non & (\Re(\nu) > 0; \mb{ and either }
|z| \leq 1;z\neq 1;\sig > 0 \mb{ or } z=1; \sig > 1)\end{align}
\begin{align}\label{eq:3.3}
\Gamma(\sigma+i\tau)\phi(z,\sigma+i\tau) & =\sqrt{2\pi}z
\mathcal{F}[\frac{e^{\sigma y}}{\exp(e^{y})-z};\tau]\non & (|z|\leq
1-\del, \del\in (0,1) \mb{ and }\sig>0;z=1\mb{ and }\sig>
1).\end{align} The BE function (1.2) can be written as
\begin{align}\label{eq:3.4} \Gamma(\sigma+i\tau)\mathfrak{B}_{\sigma+i\tau-1}(x)
=\sqrt{2\pi} e^{x}\mathcal{F}[\frac{e^{\sigma
y}}{\exp(e^{y})-e^{x}};\tau]\non(x\geq 0;\sigma>1).\end{align} The
Hurwitz and the Riemann zeta functions have the Fourier transform
representations
\begin{align}\label{eq:3.5}
\Gamma(\sigma+i\tau)\zeta(\sigma+i\tau,\nu)
=\sqrt{2\pi}\mathcal{F}[\frac{e^{\sigma y}\exp(-(\nu-1)
e^{y})}{\exp(e^{y})-1};\tau]\non (\Re(\nu)>0;\sigma>1).
\end{align}
\begin{align}\label{eq:3.6}
\Gamma(\sigma+i\tau)\zeta(\sigma+i\tau)
=\sqrt{2\pi}\mathcal{F}[\frac{e^{\sigma
y}}{\exp(e^{y})-1};\tau]\quad(\sigma>1).
\end{align} The Fourier transform representation of the eFD
function is
\begin{align}\label{eq:3.7} \Gamma(\sigma+i\tau)\Theta_\nu(\sigma+i\tau;x)
=e^{-(\nu+1) x}\sqrt{2\pi}\mathcal{F}[\frac{e^{\sigma y}\exp(-\nu
e^{y})}{\exp(e^{y})+
e^{-x}};\tau]\non\quad(\Re(\nu)>-1;x\geq0;\sigma>0).
\end{align} Similarly for the FD and the Riemann zeta
functions, we have
\begin{align}
\label{eq:3.8}\Gamma(\sigma+i\tau)\mathfrak{F}_{\sigma+i\tau-1}(x)=\sqrt{2\pi}
e^{x}\mathcal{F}[\frac{e^{\sigma
y}}{\exp(e^{y})+e^{x}};\tau]\non(x\geq0;\sigma>1).\end{align}
\begin{align}\label{eq:3.9}
C(\sigma+i\tau)\zeta(\sigma+i\tau)=\sqrt{2\pi}\mathcal{F}[\frac{e^{\sigma
y}}{\exp(e^{y})+1};\tau] \quad(\sigma>0).
\end{align}
 \section{Some applications of Fourier transform representation by using Parseval's identity} Let $f, g$ be two arbitrary Fourier transformable
functions then the Parseval's identity (\cite{aiz}, p.232) of
Fourier transform states that
\begin{align}\label{eq:4.1}\int_{-\infty}^{+\infty}\mathcal{F}[f(y);\tau]
\overline{\mathcal{F}[g(y);\tau]}d\tau &
=\int_{-\infty}^{+\infty}f(y)\overline{g(y)}dy.\end{align} Using
(3.1) and (4.1), we get the following identity
\begin{align}\label{eq:4.2}\int_{-\infty}^{+\infty}
\Gamma(\sigma+i\tau)\Gamma(\rho-i\tau)\Psi_\nu(\sigma+i\tau;x)\Psi_\nu(\rho-i\tau;x)d\tau
=2\pi\int_0^{\infty}\frac{e^{-2(\nu+1)x}e^{-2\nu
t}t^{\sigma+\rho-1}}{(e^{t}-e^{-x})^2}dt\non
(\Re(\nu)>-1;\sigma+\rho>1\mb{ when } x=0;\sigma+\rho>0\mb{ when
}x>0).\end{align} The integral on the right hand side of (4.2) can
be evaluated by replacing $\eta\mapsto\sig+\rho \mb { and }
\nu\mapsto2\nu$ in (2.10). This gives the following result for the
\emph{eBE function}
\begin{align}\label{eq:4.3} \int_{-\infty}^{+\infty}
\Gamma(\sigma+i\tau)\Gamma(\rho-i\tau)\Psi_\nu(\sigma+i\tau;x)\Psi_\nu(\rho-i\tau;x)d\tau
 ~~~~~~~~~~~~~~~~~~~~~~~~~~~~~~~~~~~~~~~\non
=2\pi\Gamma(\sigma+\rho)[\Psi_{2\nu}(\sigma+\rho-1;x)-(2\nu+1)\Psi_{2\nu}(\sigma+\rho;x)]\non
(\Re(\nu)>-1;\sigma+\rho>1\mb{ when }x>0;\sigma+\rho>2\mb{ when }
x=0)\end{align} and as a special case, taking $\rho=\sigma$, we get
\begin{align}\label{eq:4.4}\int_{-\infty}^{+\infty}
|\Gamma(\sigma+i\tau)\Psi_\nu(\sigma+i\tau;x)|^{2}d\tau  =
2\pi\Gamma(2\sigma)[\Psi_{2\nu}(2\sigma-1;x)-(2\nu+1)\Psi_{2\nu}(2\sigma;x)]\non
(\Re(\nu)>-1;\sigma>1/2\mb{ when }x>0;\sigma>1\mb{ when }
x=0).\end{align} Using (1.20) and (2.12) in (4.2), we obtain
following identity for the \emph{Hurwitz-Lerch zeta
function}\begin{align}
 \label{eq:4.5}\int_{-\infty}^{+\infty}
\Gamma(\sigma+i\tau)\Gamma(\rho-i\tau)\Phi(z,\sigma+i\tau,\nu)\Phi(z,\rho-i\tau,\nu)d\tau
~~~~~~~~~~~~~~~~~~~~~~~~~~~~~~~~~~~~~~~\non =
\frac{2\pi\Gamma(\sigma+\rho)}{
z}[\Phi(z,\sigma+\rho-1,2\nu-1)-(2\nu-1)\Phi(z,\sigma+\rho,2\nu-1)]\non(\Re(\nu)>0;\sigma+\rho>1\mb{
when }0<z<1;\sigma+\rho>2\mb{ when }z=1)
\end{align}
and as a special case, taking $\rho=\sigma$, we obtain \begin{align}
\label{eq:4.6}\int_{-\infty}^{+\infty}
|\Gamma(\sigma+i\tau)\Phi(z,\sigma+i\tau,\nu)|^{2}d\tau =
\frac{2\pi\Gamma(2\sigma)}{
z}[\Phi(z,2\sigma-1,{2\nu-1})-(2\nu-1)\Phi(z,2\sigma,{2\nu-1})]\non(\Re(\nu)>1/2;\sigma>1/2\mb{
when }0<z<1;\sigma>1\mb{ when }z=1).
\end{align}Similarly by using (1.16) and (2.13) in (4.2), we arrive at the following
result for the \emph{polylogarithm
function}\begin{align}\label{eq:4.7} \int_{-\infty} ^{+\infty}
\Gamma(\sigma+i\tau)\Gamma(\rho-i\tau)\phi(z,\sigma+i\tau)\phi(z,\rho-i\tau)d\tau=
2\pi\Gamma(\sigma+\rho)[\phi(z,\sigma+\rho-1)-\phi(z,\sigma+\rho)]\non(\sigma+\rho>1\mb{
when }0<z<1;\sigma+\rho>2\mb{ when }z=1)
\end{align}
and as a special case, taking $\rho=\sigma$, we get
\begin{align} \label{eq:4.8}\int_{-\infty}^{+\infty}
|\Gamma(\sigma+i\tau)\phi(z,\sigma+i\tau)|^{2}d\tau
=2\pi\Gamma(2\sigma)[\phi(z,2\sigma-1)-\phi(z,2\sigma)]\non(\sigma>1/2\mb{
when }0<z<1;\sigma>1\mb{ when }z=1).\end{align} Upon using (1.7) and
(2.14) in (4.2), we get the following results for the \emph{BE
function}
\begin{align} \label{eq:4.9}
\int_{-\infty}^{+\infty}\Gamma(\sigma+i\tau)\Gamma(\rho-i\tau)\mathfrak{B}_{\sigma+i\tau-1}(x)\mathfrak{B}_{\rho-i\tau-1}(x)d\tau
~~~~~~~~~~~~~~~~~~~~~~~~~~~~~~~~~~~~~~~~~~~~\non=
2\pi\Gamma(\sigma+\rho)[\mathfrak{B}_{\sigma+\rho-2}(x)-\mathfrak{B}_{\sigma+\rho-1}(x)]\quad(x\geq0;\sigma+\rho>2)\end{align}
and as a special case, taking $\rho=\sigma$, we get
\begin{align}
\label{eq:4.10}\int_{-\infty}^{+\infty}
|\Gamma(\sigma+i\tau)\mathfrak{B}_{\sigma+i\tau-1}(x)|^{2}d\tau =
2\pi\Gamma(2\sigma)[\mathfrak{B}_{2\sigma-2}(x)-\mathfrak{B}_{2\sigma-1}(x)]\quad(x\geq0;\sigma>1).\end{align}
Using (1.13) and (2.15) in (4.2), we arrive at the following
identity for the \emph{Hurwitz zeta function}
\begin{align}\label{eq:4.11}\int_{-\infty}^{+\infty}
\Gamma(\sigma+i\tau)\Gamma(\rho-i\tau)\zeta(\sigma+i\tau,\nu)\zeta(\rho-i\tau,\nu)d\tau~~~~~~~~~~~~~~~~~~~~~~~~~~~~~~~~~~~~~~~\non
=2\pi\Gamma(\sigma+\rho)[\zeta(\sigma+\rho-1,2\nu-1)-(2\nu-1)\zeta(\sigma+\rho,2\nu-1)]\non
(\Re(\nu)>1/2;\sig+\rho>2)\end{align} and as a special case, taking
$\rho=\sigma$, we get
\begin{align}\label{eq:4.12} \int_{-\infty}^{+\infty}
|\Gamma(\sigma+i\tau)\zeta(\sigma+i\tau,\nu)|^{2}d\tau=
2\pi\Gamma(2\sigma)[\zeta(2\sigma-1,{2\nu-1})-(2\nu-1)\zeta(2\sigma,{2\nu-1})]\non
(\Re(\nu)>1/2;\sig>1).
\end{align}
 Similarly by using (1.10) and (2.16) in (4.2), we arrive at the following identity for the \emph{Riemann zeta function}
\begin{align}\label{eq:4.13} \int_{-\infty}^{+\infty}
\Gamma(\sigma+i\tau)\Gamma(\rho-i\tau)\zeta(\sigma+i\tau)\zeta(\rho-i\tau)d\tau
=2\pi\Gamma(\sigma+\rho)[\zeta(\sigma+\rho-1)-\zeta(\sigma+\rho)]\non
(\sig+\rho>2)\end{align} and as a special case, taking
$\rho=\sigma$, we get
\begin{align}\label{eq:4.14}\int_{-\infty}^{+\infty}
|\Gamma(\sigma+i\tau)\zeta(\sigma+i\tau)|^{2}d\tau=
2\pi\Gamma(2\sigma)[\zeta(2\sigma-1)-\zeta(2\sigma)]\quad
(\sig>1).\end{align}

Now the Parseval's Identity for the Fourier transform representation
(3.7) of the {\it eFD function} produces the following identity
\begin{align}\label{eq:4.15}\int_{-\infty}^{+\infty}
\Gamma(\sigma+i\tau)\Gamma(\rho-i\tau)\Theta_\nu(\sigma+i\tau;x)\Theta_\nu(\rho-i\tau;x)d\tau
= 2\pi\int_0^{\infty}\frac{e^{-2(\nu+1)x}e^{-2\nu
t}t^{\sigma+\rho-1}}{(e^{t}+e^{-x})^2}dt\non
\quad(\Re(\nu)>0;x\geq0;\sig,\rho>0) .\end{align}The integral on the
right hand side of (4.15) can be evaluated by $\eta\mapsto\sig+\rho
\mb { and } \nu\mapsto2\nu$ in (2.1).  This gives following result
for the \emph{eFD function}
\begin{align}\label{eq:4.16}\int_{-\infty}^{+\infty}
\Gamma(\sigma+i\tau)\Gamma(\rho-i\tau)\Theta_\nu(\sigma+i\tau;x)\Theta_\nu(\rho-i\tau;x)d\tau
~~~~~~~~~~~~~~~~~~~~~~~~~~~~~~~~~~~~~~~\non=
2\pi\Gamma(\sigma+\rho)[(2\nu+1)\Theta_{2\nu}(\sigma+\rho;x)-\Theta_{2\nu}(\sigma+\rho-1;x)]\non(\Re(\nu>0;x\geq0;\sig+\rho>1)
\end{align}
 and as a special case, taking $\rho=\sigma$, we
get
\begin{align}\label{eq:4.17}\int_{-\infty}^{+\infty}
|\Gamma(\sigma+i\tau)\Theta_\nu(\sigma+i\tau;x)|^{2}d\tau =
2\pi\Gamma(2\sigma)[(2\nu+1)\Theta_{2\nu}(2\sigma;x)-\Theta_{2\nu}(2\sigma-1;x)]\non(\Re(\nu)>0;x\geq0;\sig>1/2).\end{align}Similarly
by using (1.6) and (2.8) in (4.16) we get the following result for
the \emph{FD function}\begin{align}
\label{eq:4.18}\int_{-\infty}^{+\infty}
\Gamma(\sigma+i\tau)\Gamma(\rho-i\tau)\mathfrak{F}_{\sigma+i\tau-1}(x)\mathfrak{F}_{\rho-i\tau-1}(x)d\tau~~~~~~~~~~~~~~~~~~~~~~~~~~~~~~~~~~~~~~~\non=
2\pi\Gamma(\sigma+\rho)[\mathfrak{F}_{\sigma+\rho-1}(x)-\mathfrak{F}_{\sigma+\rho-2}(x)]\quad(x\geq0;\sigma+\rho>1)\end{align}
and as a special case, taking $\rho=\sigma$ in (4.18), we get
\begin{align}\label{eq:4.19}\int_{-\infty}^{+\infty}
|\Gamma(\sigma+i\tau)\mathfrak{F}_{\sigma+i\tau-1}(x)|^{2}d\tau &
=2\pi\Gamma(2\sigma)[\mathfrak{F}_{2\sigma-1}(x)-\mathfrak{F}_{2\sigma-2}(x)]
\qquad(x\geq0;\sigma>1/2).\end{align} Upon using (1.11) and (2.9) in
(4.16),
  we arrive at the following identity for the \emph{Riemann zeta function}
\begin{align}\label{eq:4.20}\int_{-\infty}^{+\infty}
C(\sigma+i\tau)C(\rho-i\tau)\zeta(\sigma+i\tau)\zeta(\rho-i\tau)d\tau~~~~~~~~~~~~~~~~~~~~~~~~~~~~~~~~~~~~~~~\non
=2\pi\Gam(\sig+\rho)[(1-2^{1-\sigma-\rho})\zeta(\sigma+\rho)-(1-2^{-\sigma-\rho})\zeta(\sigma+\rho-1)]
\quad(\sigma+\rho>1).\end{align} and as a special case, taking $\rho
=\sigma$, we get
\begin{align}\label{eq:4.21}\int_{-\infty}^{+\infty}
|C(\sigma+i\tau)\zeta(\sigma+i\tau)|^{2}d\tau =
2\pi\Gam(2\sig)[(1-2^{1-2\sigma})\zeta(2\sigma)-(1-2^{-2\sigma})\zeta(2\sigma-1)]\non
(\sigma>1/2).\end{align} By making use of the Parseval's identity
for (3.2) and (3.4) we get\begin{align}
 \label{eq:4.22}\int_{-\infty}^{+\infty}
\Gamma(\sigma+i\tau)\Gamma(\rho-i\tau)\Phi(e^{-x},\sigma+i\tau,\nu)\mathfrak{B}_{\rho-i\tau-1}(-x)d\tau
~~~~~~~~~~~~~~~~~~~~~~~~~~~~~~~~~~~~~~~\non =
2\pi\Gamma(\sigma+\rho)[\Phi(e^{-x},\sigma+\rho-1,\nu)-\nu\Phi(e^{-x},\sigma+\rho,\nu)]\non(\Re(\nu)>0;\sigma+\rho>1\mb{
when }x>0;\sigma+\rho>2\mb{ when }x=0)
\end{align}Here the left hand side is obtained by using (2.12),
however, for $\nu=1$ in the above relation we can obtain (4.9).
 Now by making use of (2.14) and the Parseval's identity for (3.3) and (3.4), we get
\begin{align}
 \label{eq:4.23}\int_{-\infty}^{+\infty}
\Gamma(\sigma+i\tau)\Gamma(\rho-i\tau)\mathfrak{B}_{\sig+i\tau-1}(-x)\phi(e^{-x},\rho-i\tau)d\tau
~~~~~~~~~~~~~~~~~~~~~~~~~~~~~~~~~~~~~~~\non =
2\pi\Gamma(\sigma+\rho)[\mathfrak{B}_{\sigma+\rho-2}(-x)-\mathfrak{B}_{\sigma+\rho-1}(-x))]\non(\sigma+\rho>1\mb{
when }x>0;\sigma+\rho>2\mb{ when }x=0)
\end{align} By making use of (2.13) in place of (2.14), the left hand side of above equation can also be written in terms of polylogarirhm function.
By making use of (2.15) and the Parseval's identity for (3.5) and
(3.6), we get\begin{align}\label{eq:4.24}\int_{-\infty}^{+\infty}
\Gamma(\sigma+i\tau)\Gamma(\rho-i\tau)\zeta(\sigma+i\tau,\nu)\zeta(\rho-i\tau)d\tau~~~~~~~~~~~~~~~~~~~~~~~~~~~~~~~~~~~~~~~\non
=2\pi\Gamma(\sigma+\rho)[\zeta(\sigma+\rho-1,\nu)-\nu\zeta(\sigma+\rho,\nu)]\non
(\Re(\nu)>0;\sig+\rho>2)\end{align}For $\nu=1$ the above identity
reduces to (4.13). Now by applying Parseval's identity for (3.7) and
(3.8), we get\begin{align}\label{eq:4.25}\int_{-\infty}^{+\infty}
\Gamma(\sigma+i\tau)\Gamma(\rho-i\tau)\Theta_\nu(\sigma+i\tau;x)\mathfrak{F}_{\rho-i\tau-1}(-x)d\tau
~~~~~~~~~~~~~~~~~~~~~~~~~~~~~~~~~~~~~~~\non=
2\pi\Gamma(\sigma+\rho)[(\nu+1)\Theta_{\nu}(\sigma+\rho;x)-\Theta_{\nu}(\sigma+\rho-1;x)]\non(\Re(\nu)>-1;x\geq0;\sig+\rho>1).
\end{align} However for $\nu=0$, this reduces to (4.18). Similarly
more results can be obtained by making use of the Parseval's
identity for Fourier transform representations of different
functions obtained in section 3.
\section{Some applications of Fourier transform representation by using duality property} The duality property of Fourier transform (\cite{ld}, p. 29)
states that \bqn\label{eq:5.1}
\mathcal{F}[\mathcal{F}[\varphi(y);\tau];\omega]=\varphi(-\omega),\quad\omega\in\R.\9
This equation gives Fourier transform of the \emph{eBE} function
when we take Fourier transform on the both sides of (3.1)
\begin{align}\label{eq:5.2}\mathcal{F}[\Gamma(\sigma+i\tau)\Psi_\nu(\sigma+i\tau;x);\omega] & =\frac{\sqrt{2\pi} e^{-(\nu+1)x}e^{-\sigma \omega}\exp(-\nu
e^{-\omega})}{\exp(e^{-\omega})-e^{-x}}\non &
(\Re(\nu)>-1;\sig>0\mb{ when }x> 0;\sig>1\mb{ when }x=0).
\end{align} Similarly for the \emph{eFD} function we have
\begin{align}\label{eq:5.3}\mathcal{F}[\Gamma(\sigma+i\tau)\Theta_\nu(\sigma+i\tau;x);\omega] & =
\frac{\sqrt{2\pi} e^{-(\nu+1)x}e^{-\sigma \omega}\exp(-\nu
e^{-\omega})}{\exp(e^{-\omega})+e^{-x}}\qquad(\Re(\nu)>-1;x\geq
0;\sig>0).
\end{align}
 For the \emph{Hurwitz-Lerch zeta} function
\begin{align}\label{eq:5.4}
\mathcal{F}[\Gam(\sig+i\tau)\Phi(z,\sigma+i\tau,\nu);\omega] & =
\frac{\sqrt{2\pi} e^{-\sigma \omega}\exp(-(\nu-1)
e^{-\omega})}{\exp(e^{-\omega})-z}\non & (\Re(\nu) > 0; \mb{ and
either } |z| \leq 1;z\neq 1;\sig > 0 \mb{ or } z=1; \sig >
1).\end{align}Similarly  for the \emph{Polylogarithm} function
\label{eq:5.5}\begin{align}
\mathcal{F}[\Gamma(\sigma+i\tau)\phi(\sigma+i\tau,z);\omega] &
=\frac{\sqrt{2\pi} ze^{-\sig \omega}} {\exp(e^{-\omega})-z}\non &
(|z|\leq 1-\del, \del\in (0,1) \mb{ and }\sig>0;z=1\mb{ and }\sig>
1).\end{align} For the \emph{BE} function
\begin{align}\label{eq:5.6}
\mathcal{F}[\Gamma(\sigma+i\tau)\mathfrak{B}_{\sigma+i\tau-1}(x);\omega]
& =\frac{\sqrt{2\pi} e^{x}e^{-\sig \omega}}
{\exp(e^{-\omega})-e^{x}}\qquad(x\geq0;\sig>1).\end{align}Similarly
 for the \emph{FD function}
 \begin{align} \label{eq:5.7}
\mathcal{F}[\Gamma(\sigma+i\tau)\mathfrak{F}_{\sigma+i\tau-1}(x);\omega]
& =\frac{\sqrt{2\pi} e^{x}e^{-\sig \omega}}
{\exp(e^{-\omega})+e^{x}} \qquad(x\geq0;\sig>0).\end{align} We get
following expression involving the \emph{Hurwitz zeta function}

\begin{align} \label{eq:5.8}
\mathcal{F}[\Gamma(\sigma+i\tau)\zeta(\sigma+i\tau,\nu);\omega]=\frac{\sqrt{2\pi}
e^{-\sig \omega}\exp(-(\nu-1)e^{-\omega})}{\exp(e^{-\omega})-1}\quad
(\Re(\nu)>0;\sig>1).\end{align}Similarly we get following formulae
for the \emph{Riemann zeta} function
\begin{align}\label{eq:5.9}
\mathcal{F}[\Gam(\sigma+i\tau)\zeta(\sigma+i\tau);\omega]=\frac{\sqrt{2\pi}
e^{-\sigma \omega}}{\exp(e^{-\omega})-1} \quad(\sigma>1).
\end{align}
\begin{align}\label{eq:5.10}
\mathcal{F}[C(\sigma+i\tau)\zeta(\sigma+i\tau);\omega]=\frac{\sqrt{2\pi}
e^{-\sigma \omega}}{\exp(e^{-\omega})+1} \quad(\sigma>0).
\end{align}
\brem \h2 {\bf 1.} We have already obtained the Fourier transform of
the Riemann zeta function for $\sig>0$. It would be more interesting
to have the Fourier transform of the Riemann zeta function in the
critical strip. The Riemann zeta function has the Mellin transform
representation (\cite{cz}, p. 294, Equation
(7.48))\begin{align}\label{eq:5.11} \zeta(s)= \fr{1}{\Gam(s)}
\int^\inf_0 t^{s-1}\big(\fr{1}{e^t-1}-\fr{1}{t})dt\qquad
(0<\sig<1).\end{align} Again the substitution $t=e^{y}$ in (5.11)
yields the Fourier transform representation
\begin{align}\label{eq:5.12}
\Gamma(\sigma+i\tau)\zeta(\sigma+i\tau)=\sqrt{2\pi}\mathcal{F}[e^{\sigma
y}[\frac{1}{\exp(e^{y})-1}-\frac{1}{e^{y}}];\tau]\non
\qquad(0<\sigma<1).
\end{align}
This gives\begin{align}\label{eq:5.13}
\mathcal{F}[\Gamma(\sigma+i\tau)\zeta(\sigma+i\tau);\omega] &
=\frac{ \sqrt{2\pi} e^{-\sig
\omega}}{\exp(e^{-\omega})-1}-\sqrt{2\pi} e^{(1-\sig) \omega} &
(0<\sigma<1).\end{align}\erem \brem \h2 {\bf 2.} We note that
special cases of these Fourier transforms give following interesting
integral formulas. For $\omega=0$, we have
\begin{align}\label{eq:5.14}\int_{-\infty}^{+\infty}\Gamma(\sigma+i\tau)\Psi_\nu(\sigma+i\tau;x)d\tau & =\frac{2\pi
e^{-\nu(x+1)}}{e^{x+1}-1}\non & (\Re(\nu)>-1;\sig>0\mb{ when }x>
0;\sig>1\mb{ when }x=0).
\end{align}\begin{align}\label{eq:5.15}\int_{-\infty}^{+\infty}\Gamma(\sigma+i\tau)\Theta_\nu(\sigma+i\tau;x)d\tau & =\frac{2\pi
e^{-\nu(x+1)}}{e^{x+1}+1}\non & (\Re(\nu)>-1;\sig>0\mb{ when }x>
0;\sig>1\mb{ when }x=0).
\end{align}\begin{align}\label{eq:5.16}\int_{-\infty}^{+\infty}\Gamma(\sigma+i\tau)\Phi(\sigma+i\tau;z,\nu)d\tau & =\frac{2\pi
e^{1-\nu}}{e-z}\non & (\Re(\nu) > 0; \mb{ and either } |z| \leq
1;z\neq 1;\sig > 0 \mb{ or } z=1; \sig > 1).
\end{align}\begin{align}\label{eq:5.17} \int_{-\infty}^{+\infty}{\Gamma(\sigma+i\tau)}\phi(z,\sigma+i\tau)
& =\frac{2\pi z}{e-z}\non & (|z|\leq 1-\del, \del\in (0,1) \mb{ and
}\sig>0;z=1\mb{ and }\sig>
1).\end{align}\begin{align}\label{eq:5.18}
\int_{-\infty}^{+\infty}\Gamma(\sigma+i\tau)\mathfrak{B}_{\sigma+i\tau-1}(x)d\tau
& =\frac{2\pi}{e^{1-x}-1}\qquad(x\geq0;\sig>1).\end{align}
 \begin{align} \label{eq:5.19}
\int_{-\infty}^{+\infty}\Gamma(\sigma+i\tau)\mathfrak{F}_{\sigma+i\tau-1}(x)d\tau
& =\frac{2\pi} {e^{1-x}+1} \qquad(x\geq0;\sig>0).\end{align}
\begin{align} \label{eq:5.20}
\int_{-\infty}^{+\infty}\Gamma(\sigma+i\tau)\zeta(\sigma+i\tau,\nu)d\tau=\frac{2\pi
e^{1-\nu}}{e-1}\quad (\Re(\nu)>0;\sig>1).\end{align}
\begin{align}\label{eq:5.21}
\int_{-\infty}^{+\infty}C(\sigma+i\tau)\zeta(\sigma+i\tau)d\tau=\frac{2\pi}{e+1}
\quad(\sigma>0).
\end{align}
\bqn\label{eq:5.22}
\int_{-\infty}^{+\infty}\Gamma(\sigma+i\tau)\zeta(\sigma+i\tau)d\tau=\fr{2\pi}{e-1}\quad(\sig>1).\9
\begin{align}\label{eq:5.23}
\int_{-\infty}^{+\infty}\Gam(\sigma+i\tau)\zeta(\sigma+i\tau)d\tau=\frac{2\pi}{e-1}-2\pi
e \quad(0<\sigma<1).
\end{align}
\erem

\section{Concluding remarks}

In this paper we have obtained the Fourier transform representation
of the eFD and eBE functions defined in \cite{scqt}. These functions
have elegant connections with the zeta family and other related
functions. These connections have been used to get new integral
formulae for these functions.

In section 2 two identities involving the eFD and eBE functions were
proved. This led to new results involving the FD, BE, polylogarithm
and Hurwitz-Lerch zeta functions. Results for Riemann and Hurwitz
zeta functions have been deduced as special cases. Parseval's
identity of the Fourier transform proved crucial in obtaining the
functional relations of the integrals of the product of all these
functions with the gamma function, by using these new identities.
However, these identities can be used further to get inequalities
involving the eFD and eBE functions, which will prove useful to get
inequalities for the Hurwitz-Lerch zeta, FD and BE functions. This
will be discussed in a future paper \cite{ataq}.

The classical theory of Fourier transforms provides a number of
mathematical tools that could be useful in solving many problems of
interest. Here we
have used the duality property of the Fourier transform to find the Fourier
transforms of the product of the gamma function with the higher
transcendental functions related to the zeta family. It leads to some
interesting special cases, which give integral  formulae involving
zeta related functions. This approach will hopefully enhance the
applicability of these functions in various physical and engineering
problems.

Most of the interest in the study of the Riemann zeta function comes
from the critical strip because of the famous unproved Riemann
hypothesis. We have provided the Fourier transform representation of
the Riemann zeta function in the critical strip. This proved useful
to evaluate the Fourier transform of the Riemann zeta function in
the critical strip. It gives new insights that other special
functions having Mellin and hence Fourier transform representations
can be used to get more integral formulae. For example by using
Parseval's identity for the Fourier transform representations of the
Riemann zeta function (3.6) and the Euler gamma function (\cite{tq},
p. 3), we get\begin{align}\label{eq:6.1}
\int_{-\infty}^{+\infty}\Gam(\rho-i\tau)\Gam(\sigma+i\tau)\zeta(\sigma+i\tau)d\tau=\z(\sig+\rho,2)\quad(\rho,\sigma>1).
\end{align}Special functions satisfy certain recurrence relations, which can also be used to
evaluate more integral formulae. For example the eFD function
satisfies (\cite{scqt},p. 15, Equation 5.19)\bqn
\label{eq:6.2}\Theta_\nu(s;x) + \Theta_{\nu-1}(s;x) =
\nu^{-s}e^{-\nu x} \quad (x \geq 0, \nu\geq 1), \9 which along with
(5.15) gives
\begin{align}\label{eq:6.3}
\int_{-\infty}^{+\infty}{\nu}^{-\sig-i\tau}\Gam(\sigma+i\tau)d\tau=2\pi
e^{-\nu}\quad(\nu\geq 1,\sigma>0).
\end{align} It is hoped that other properties of Fourier transform
will be useful to get more results involving these extended and
other related functions.

\section*{Acknowledgements}

AT acknowledges her indebtedness to the Higher Education Commission
of the Government of Pakistan for the Indigenous Ph.D. Fellowship.

\end{document}